\documentclass[%
reprint,
amsmath,amssymb,
aps,
pra,
floatfix,
]{revtex4-2}

\usepackage[]{graphicx}
\usepackage{dcolumn}
\usepackage{bm}

\begin{document}
\setlength\textfloatsep{6pt}


\title{Insensitivity of the two-photon Jaynes-Cummings model to thermal noise}

\author{Hiroo Azuma}
\email{zuma@nii.ac.jp}
\affiliation{Global Research Center for Quantum Information Science,
National Institute of Informatics, 2-1-2 Hitotsubashi, Chiyoda-ku, Tokyo 101-8430, Japan}




\date{\today}

\begin{abstract}
We study the thermal effects of the multiphoton Jaynes-Cummings model (JCM) using a thermofield dynamics (TFD) method.
Letting the initial state of the whole system for the multiphoton JCM be a product of the ground state of an atom and a coherent state of a cavity field
at finite temperature,
we compute its time evolution.
We evaluate a period of the collapse and revival of the Rabi oscillations and the relative entropy of coherence of the atom
up to the second-order perturbation of the low-temperature expansion.
We show that an intuitive estimation of the period matches the result of the perturbation theory of TFD well.
In particular, we see that the period of the two-photon JCM hardly depends
on the amplitude of the coherent state of the cavity field or the temperature.
Numerical calculations suggest that the relative entropy of coherence of the two-photon JCM does not decay even for nonzero-temperature cases as time proceeds.
By contrast, the relative entropyies of coherence for single-, three-, and four-photon JCMs decay as time proceeds for zero- and finite-temperature cases.
\end{abstract}

\maketitle

\section{\label{section-introduction}Introduction}
The Jaynes-Cummings model (JCM) is theoretically and experimentally successful in the field of quantum optics.
The JCM was first proposed by Jaynes and Cummings in 1963 \cite{Jaynes1963}.
It is a soluble fully quantum mechanical model that describes the spontaneous emission of a cavity field interacting with an atom \cite{Louisell1973,Shore1993,Schleich2001}.
One of the remarkable characteristics of the JCM is the collapse and revival of Rabi oscillations during its time evolution \cite{Eberly1980,Narozhny1981,Yoo1981,Yoo1985}.
This phenomenon was demonstrated experimentally \cite{Rempe1987},
and it is regarded as direct evidence of the statistical and discrete nature of the quantum field of photons, which has no classical counterpart.

A multiphoton JCM is a natural extension of the JCM.
In the ordinary JCM, the single annihilation and creation operators of the cavity field couple with the atomic raising and lowering operators, respectively.
In contrast, the multiphoton JCM requires a multiphoton transition process which allows coupling
between multiples of annihilation and creation operators of photons and the raising and lowering operators of the atom,
respectively.

The multiphoton JCM is attractive not just from a theoretical context; some researchers think of it as a realistic candidate for experiments.
Implementation of the two-photon JCM with a superconducting quantum interference device (SQUID) was proposed \cite{Felicetti2018a}.
It was shown that the nonlinear interaction of the SQUID between a flux qubit and a bosonic mode leads to a nondipolar term and it causes the two-photon quantum Rabi model (QRM).
Then, simplifying the two-photon QRM with the rotating-wave approximation,
we obtain the two-photon JCM.
Implementation of the two-photon QRM was also proposed for trapped ions \cite{Felicetti2015,Puebla2017}.
Thus, we can construct the two-photon JCM with trapped ions.
A scheme to realize the two-photon JCM with the interplay between laser detuning and a cavity-driven field,
two-photon bundles, and photon-induced tunneling was proposed \cite{Tang2023}.
Multiphoton quantum Rabi oscillations in ultrastrong cavity QED were studied
\cite{Garziano2015},
and they contribute to implementing the multiphoton QRM and the multiphoton JCM.
A circuit-QED scheme for realizing the ultrastrong-coupling regime of nondipolar light-matter interactions and the two-photon QRM was investigated
\cite{Felicetti2018b}.
Theoretical aspects of the multiphoton JCM were studied in
\cite{Joshi1998,Abdel-Aty2002,Tan2011,Mojaveri2018,Zou2020,Fakhri2021}.

Recently, the interaction of the two-photon coupling has attracted attention
in the field of quantum thermodynamics for applications of quantum batteries
\cite{Crescente2020,Delmonte2021}.
In \cite{Piccione2022}, a damped quantum harmonic oscillator interacting with $N$ two-level systems via two-photon coupling was studied.
Its Hamiltonian was similar to the two-photon JCM.
That work supposed that the harmonic oscillator and two-level systems were in contact with independent heat baths, and the dissipation was investigated with the master equation.

The relative entropy of coherence was proposed to quantify the coherence of an arbitrary quantum state
\cite{Baumgratz2014}.
Thus, the relative entropy of coherence can be an indicator of quantumness for the state of the system of interest,
that is, how far the state is from its classical version.
Explicitly, the relative entropy of coherence is defined as
\begin{equation}
C_{\mbox{\scriptsize rel.ent.}}(\rho)
=
S(\rho_{\mbox{\scriptsize diag}})-S(\rho),
\label{definition-relative-entropy-coherence-0}
\end{equation}
where $S$ and $\rho$ represent the von Neumann entropy and the density matrix of the system, respectively.
A diagonal matrix $\rho_{\mbox{\scriptsize diag}}$ is obtained from $\rho$ by deleting all off-diagonal elements.
An operational interpretation was given for the relative entropy of coherence \cite{Winter2016}.
As a quantity similar to the relative entropy of coherence,
the maximum relative entropy of coherence was introduced \cite{Bu2017}.

Thermofield dynamics (TFD) is a method for deriving the physical quantities
of isolated and/or closed systems
under thermal equilibrium
\cite{Takahashi1975,Umezawa1982,Umezawa1992}.
TFD introduces a fictional Hilbert space beside the original Hilbert space where the physical system is defined and assumes the two-mode squeezed vacuum state.
By tracing out the degrees of freedom of the fictional Hilbert space,
we can obtain the genuine density matrix for the original system.

The thermal effects of the period of the collapse and revival of the Rabi oscillations for the original JCM at low temperatures were evaluated with TFD \cite{Azuma2011}.
The time variation of the relative entropy of coherence for the Bixon-Jortner model at zero temperature was studied \cite{Azuma2018}.

In this paper, we compute the period of the collapse and revival of the Rabi oscillations
and the relative entropy of coherence for the multiphoton JCM at low temperatures.
According to TFD,
we calculate them up to the second-order perturbation of the low-temperature expansion.
First, we derive the zero-temperature period of the collapse and revival of the Rabi oscillations for the multiphoton JCM, whose interaction term is given by
$\sigma_{+}a^{l}+\sigma_{-}(a^{\dagger})^{l}$,
where $\sigma_{+}$ and $\sigma_{-}$ are atomic raising and lowering operators, respectively,
$a$ and $a^{\dagger}$ are annihilation and creation operators of the cavity field, respectively,
and $l=1,2,3,...$.
Second, we estimate the period approximately with an intuitive method.
Third, we compute the period and the relative entropy of coherence up to the second order of low-temperature expansion with TFD.
Fourth, carrying out numerical calculations, we show that the period obtained with an intuitive estimation
can be a good approximation of the period derived from the second-order perturbation theory of TFD.
Numerical results suggest that the relative entropies of coherence for the single-, three-, and four-photon JCMs decay as time proceeds in the zero- and finite-temperature cases.
Contrastingly,
the relative entropy of coherence for the two-photon JCM does not decay as time proceeds at zero or finite temperature.

One of the results obtained in this paper is as follows.
For the two-photon JCM, the period of the collapse and revival of the Rabi oscillations hardly depends on the amplitude of the coherent state of the cavity field.
Moreover, it scarcely suffers from thermal effects.
Thus, we can determine the coupling constant of interaction between the photons and atom for the two-photon JCM with experimental measurements with ease.
Moreover, as mentioned above, the relative entropy of coherence of the two-photon JCM is not affected by thermal noise.
Thus, we can regard the behavior of the two-photon JCM with an initial coherent state of the cavity field as being insensitive to thermal effects.

This paper is organized as follows.
In Sec.~\ref{section-collapse-revival-zero-temperature},
we compute the period of the collapse and revival of the Rabi oscillations for the multiphoton JCM at zero temperature.
In Secs.~\ref{section-perturbative-calculation-period} and \ref{section-perturbative-calculation-relative-entropy},
we derive the thermal effects of the period and the relative entropy of coherence with the second-order perturbation theory of TFD, respectively.
In Sec.~\ref{section-numerical-calculations}, we carry out numerical calculations of the period and the relative entropy of coherence for low temperatures.
In Sec.~\ref{section-discussion}, we provide a discussion.
In Appendixes~\ref{section-derivation-equation-15}, \ref{section-perturbative-expansion-Pe},
and \ref{section-perturbative-expansion-rho-01},
we show derivations of the mathematical expressions of the equations.
In Appendix~\ref{section-review-TFD}, we provide a brief review of TFD.

\section{\label{section-collapse-revival-zero-temperature}The period of the collapse and revival of the Rabi oscillations for the multiphoton JCM
at zero temperature}
The Hamiltonian of the multiphoton JCM is given by
\begin{equation}
H
=
\frac{\omega_{0}}{2}\sigma_{z}+\omega a^{\dagger}a+g[\sigma_{+}a^{l}+\sigma_{-}(a^{\dagger})^{l}],
\end{equation}
where we set $\hbar=1$, the angular frequency $\omega_{0}$ corresponds
to the difference in energies of the excited and ground states of the atom,
$\omega$ represents the angular frequency of the cavity field,
$g$ denotes the coupling constant,
and $l=1,2,3,...$.
Moreover, the annihilation and creation operators of the cavity field,
$a$ and $a^{\dagger}$, have the commutator $[a,a^{\dagger}]=1$,
and we write the Pauli matrix and the lowering and raising operators of the atom in the forms
\begin{equation}
\sigma_{z}
=
\left(
\begin{array}{cc}
1 & 0 \\
0 & -1
\end{array}
\right),
\quad
\sigma_{-}
=
\left(
\begin{array}{cc}
0 & 0 \\
1 & 0
\end{array}
\right),
\quad
\sigma_{+}
=
\left(
\begin{array}{cc}
0 & 1 \\
0 & 0
\end{array}
\right),
\end{equation}
where the ground and excited states of the atom are given by
$|g\rangle=(0,1)^{\mbox{\scriptsize T}}$
and
$|e\rangle=(1,0)^{\mbox{\scriptsize T}}$, respectively.
To introduce the interaction picture, we divide the Hamiltonian $H$ as
\begin{equation}
H=C_{1}+C_{2},
\end{equation}
\begin{equation}
C_{1}=\omega[(l/2)\sigma_{z}+a^{\dagger}a],
\end{equation}
\begin{equation}
C_{2}=-(\Delta/2)\sigma_{z}+g[\sigma_{+}a^{l}+\sigma_{-}(a^{\dagger})^{l}],
\end{equation}
\begin{equation}
\Delta=-\omega_{0}+l\omega.
\label{definition-Delta-0}
\end{equation}
Because $[C_{1},C_{2}]=0$,
we can describe a wave function of the system with the interaction picture as
\begin{equation}
|\Psi_{\mbox{\scriptsize I}}(t)\rangle=U(t)|\Psi_{\mbox{\scriptsize I}}(0)\rangle,
\end{equation}
where
\begin{eqnarray}
U(t)
&=&
\exp(-iC_{2}t) \nonumber \\
&=&
\left(
\begin{array}{cc}
u_{00} & u_{01} \\
u_{10} & u_{11}
\end{array}
\right),
\label{definition-U-t-0}
\end{eqnarray}
\begin{eqnarray}
u_{00}
&=&
\cos(\sqrt{D}t)+i\frac{\Delta}{2}\frac{\sin(\sqrt{D}t)}{\sqrt{D}}, \nonumber \\
u_{01}
&=&
-ig\frac{\sin(\sqrt{D}t)}{\sqrt{D}}a^{l}, \nonumber \\
u_{10}
&=&
-ig\frac{\sin(\sqrt{D'}t)}{\sqrt{D'}}(a^{\dagger})^{l}, \nonumber \\
u_{11}
&=&
\cos(\sqrt{D'}t)-i\frac{\Delta}{2}\frac{\sin(\sqrt{D'}t)}{\sqrt{D'}},
\label{matrix-elements-U-0}
\end{eqnarray}
\begin{eqnarray}
D
&=&
(\Delta/2)^{2}+g^{2}a^{l}(a^{\dagger})^{l}, \nonumber \\
D'
&=&
(\Delta/2)^{2}+g^{2}(a^{\dagger})^{l}a^{l}.
\label{explicit-forms-D-D-prime-operators}
\end{eqnarray}

Here, we set the initial state of the atom and cavity photons as
$|\Psi_{\mbox{\scriptsize I}}(0)\rangle=|g\rangle_{\mbox{\scriptsize A}}|\alpha\rangle_{\mbox{\scriptsize P}}$,
where $|\alpha\rangle_{\mbox{\scriptsize P}}$ denotes a coherent state.
Then, the probability that we observe the excited state of the atom is given by
\begin{eqnarray}
P_{\mbox{\scriptsize e}}(t)
&=&
|_{\mbox{\scriptsize A}}\langle e|\Psi_{\mbox{\scriptsize I}}(t)\rangle|^{2} \nonumber \\
&=&
g^{2}|\alpha|^{2l}e^{-|\alpha|^{2}}
\sum_{m=0}^{\infty}
\frac{|\alpha|^{2m}}{m!}
\frac{\sin^{2}(\sqrt{D_{m}}t)}{D_{m}},
\label{Prob-e-t-0}
\end{eqnarray}
where $D_{m}=(\Delta/2)^{2}+g^{2}\prod_{k=1}^{l}(m+k)$.
In the derivation of Eq.~(\ref{Prob-e-t-0}), we use $D|n\rangle_{\mbox{\scriptsize P}}=D_{n}|n\rangle_{\mbox{\scriptsize P}}$,
where $|n\rangle_{\mbox{\scriptsize P}}$ represents the number state of the photons.

Next, we estimate a period of the collapse and revival of the Rabi oscillations.
Looking at Eq.~(\ref{Prob-e-t-0}), we note that $P_{\mbox{\scriptsize e}}(t)$ is similar to the Poisson distribution
$|\alpha|^{2k}e^{-|\alpha|^{2}}/k!$.
Thus, major contributions are terms of $m\simeq |\alpha|^{2}$.
Here, for the sake of simplicity, we assume $\Delta=0$ and $|\alpha|\gg 1$.
Then, we obtain $D_{m}\simeq g^{2}m^{l}$ because $m\simeq |\alpha|^{2}\gg k$ for $k=1,...,l$.
Hence, the approximate form of $P_{\mbox{\scriptsize e}}(t)$ is given by
\begin{eqnarray}
P_{\mbox{\scriptsize e}}(t)
&\simeq&
|\alpha|^{2l}e^{-|\alpha|^{2}}\sum_{m=0}^{\infty}\frac{|\alpha|^{2m}}{m! m^{l}}\sin^{2}(gm^{l/2}t) \nonumber \\
&\simeq&
\frac{1}{2}
-
\frac{1}{2}
e^{-|\alpha|^{2}}\sum_{m=0}^{\infty}\frac{|\alpha|^{2m}}{m!}
\cos(2gm^{l/2}t),
\label{Pe-approximation-form-0}
\end{eqnarray}
where we use $m^{l}\simeq|\alpha|^{2l}$.

Because $m\simeq|\alpha|^{2}$
and
$\sqrt{m}\simeq(|\alpha|^{2}+m)/(2|\alpha|)$, we can derive the following relationships:
\begin{equation}
m^{l/2}
\simeq
|\alpha|^{l-2}(|\alpha|^{2}+lm)/2.
\label{approx-m_to-l-over-2}
\end{equation}
Then, we obtain
\begin{eqnarray}
&&
\sum_{m=0}^{\infty}
\frac{|\alpha|^{2m}}{m!}\cos(2gm^{l/2}t) \nonumber \\
&\simeq &
\exp[|\alpha|^{2}\cos(g|\alpha|^{l-2}lt)] \nonumber \\
&&
\times
\cos[g|\alpha|^{l}t+|\alpha|^{2}\sin(g|\alpha|^{l-2}lt)],
\label{approximation-derivation-Pe}
\end{eqnarray}
whose derivation is given in Appendix~\ref{section-derivation-equation-15}.
In the above equation, the term $\cos[g|\alpha|^{l}t+|\alpha|^{2}\sin(2|\alpha|^{l-2}lt)]$ causes the Rabi oscillation.
Here, we consider the special case of $l=1$.
Then, the Rabi oscillation is induced by the function
$\cos[g|\alpha|t+|\alpha|^{2}\sin(gt/|\alpha|)]$.
If we assume $|\alpha|\gg 1$ and $0<gt\ll 1$,
this function approximates to $\cos(2g|\alpha|t)$, and its period is equal to
\begin{equation}
\tau_{1}=\frac{\pi}{g|\alpha|}.
\label{period-Rabi-oscillations-l-1}
\end{equation}
Furthermore, the term $\exp[|\alpha|^{2}\cos(g|\alpha|^{l-2}lt)]$ gives rise to the collapse and revival of the Rabi oscillations,
and its period is given by
\begin{equation}
T_{0}(l)=
\frac{2\pi}{g|\alpha|^{l-2}l}.
\label{period-collapses-revivals-Rabi-oscillations-0}
\end{equation}
Here, we pay attention to the following facts.
When $l=2$,
we obtain $T_{0}(2)=\pi/g$, and it does not depend on the amplitude of the coherent light $|\alpha|^{2}$.

\section{\label{section-perturbative-calculation-period}Perturbative calculations of thermal effects of the period}
In this section, according to TFD,
we compute the thermal effects of the period of the multiphoton JCM.
A brief review of TFD is given in Appendix~\ref{section-review-TFD},
where the notations of this section are explained.
First, we define the Hamiltonians of the multiphoton JCM on ${\cal H}$ and $\tilde{\cal H}$ as
\begin{eqnarray}
\hat{H}
&=&
H-\tilde{H}, \nonumber \\
H
&=&
\frac{\omega_{0}}{2}(2c^{\dagger}c-1)+\omega a^{\dagger}a+g[c^{\dagger}a^{l}+c(a^{\dagger})^{l}], \nonumber \\
\tilde{H}
&=&
\frac{\omega_{0}}{2}(2\tilde{c}^{\dagger}\tilde{c}-1)
+\omega\tilde{a}^{\dagger}\tilde{a}
+g[\tilde{c}^{\dagger}\tilde{a}^{l}+\tilde{c}(\tilde{a}^{\dagger})^{l}],
\end{eqnarray}
where we put a hat symbol on the total Hamiltonian, $\hat{H}$.
The Hamiltonian $\tilde{H}$ is defined on the fictional Hilbert space.

Next, we divide $\hat{H}$ as follows:
\begin{eqnarray}
\hat{H}
&=&
\hat{C}_{1}+\hat{C}_{2}, \nonumber \\
\hat{C}_{1}
&=&
\omega[l(c^{\dagger}c-\tilde{c}^{\dagger}\tilde{c})+(a^{\dagger}a-\tilde{a}^{\dagger}\tilde{a})], \nonumber \\
\hat{C}_{2}
&=&
g[c^{\dagger}a^{l}+c(a^{\dagger})^{l}-\tilde{c}^{\dagger}\tilde{a}^{l}-\tilde{c}(\tilde{a}^{\dagger})^{l}] \nonumber \\
&&
-\Delta(c^{\dagger}c-\tilde{c}^{\dagger}\tilde{c}),
\end{eqnarray}
where $\Delta$ is given by Eq.~(\ref{definition-Delta-0}).
Because $[\hat{C}_{1},\hat{C}_{2}]=0$, the unitary time-evolution operator is given by
\begin{eqnarray}
\hat{U}(t)
&=&
\exp(-i\hat{C}_{2}t)
=
U(t)\otimes\tilde{U}(t), \nonumber \\
U(t)
&=&
\exp
\left[
-it
\left(
\begin{array}{cc}
-\Delta/2 & ga^{l} \\
g(a^{\dagger})^{l} & \Delta/2
\end{array}
\right)
\right], \nonumber \\
\tilde{U}(t)
&=&
\exp
\left[
it
\left(
\begin{array}{cc}
-\Delta/2 & g\tilde{a}^{l} \\
g(\tilde{a}^{\dagger})^{l} & \Delta/2
\end{array}
\right)
\right].
\label{definition-U-hat-t-0}
\end{eqnarray}

Before we get into the rigorous calculations for the perturbative expansion,
we make an intuitive estimation of the period of the collapse and revival of the Rabi oscillations at finite temperatures.
The period for zero temperature is given by Eq.~(\ref{period-collapses-revivals-Rabi-oscillations-0}).
Here, we evaluate $|\alpha|^{2}$ at finite temperature.
Because the square of the absolute value of the amplitude is described as $|\alpha|^{2}=\langle\alpha|a^{\dagger}a|\alpha\rangle$,
we compute its thermalized value as $\langle\alpha;\theta|a^{\dagger}a|\alpha;\theta\rangle$,
where $|\alpha;\theta\rangle$ is a thermal coherent state given
by Eq.~(\ref{definition-thermal-coherent-state-1}). 
Thus,
we obtain
\begin{eqnarray}
&&
\langle\alpha;\theta|a^{\dagger}a|\alpha;\theta\rangle \nonumber \\
&=&
{}_{\mbox{\scriptsize B}}\langle\alpha|_{\tilde{\mbox{\scriptsize B}}}\langle\alpha^{*}|
a^{\dagger}(-\theta)a(-\theta)
|\alpha\rangle_{\mbox{\scriptsize B}}|\alpha^{*}\rangle_{\tilde{\mbox{\scriptsize B}}} \nonumber \\
&=&
|\alpha|^{2}[1+2\theta(\beta)+2\theta(\beta)^{2}]
+
\theta(\beta)^{2}+{\cal O}[\theta(\beta)^{3}],
\end{eqnarray}
where $\theta(\beta)$ and $a(-\theta)$ are defined in Eqs.~(\ref{definition-theta-beta}) and (\ref{definition-a-theta}), respectively.
Because of the thermal effects,
assuming $|\alpha|\gg 1$ and $\theta(\beta)\ll 1$,
we can expect that the period
$T_{0}(l)$ given by Eq.~(\ref{period-collapses-revivals-Rabi-oscillations-0})
changes into
\begin{equation}
T'_{0}(l)
\simeq
\frac{2\pi}{gl}
\{
|\alpha|^{2}
[1+2\theta(\beta)+2\theta(\beta)^{2}]
+
\theta(\beta)^{2}
\}
^{1-(l/2)}.
\label{Period-beta-dependence-approximate-1}
\end{equation}
Here, setting $k_{\mbox{\scriptsize B}}=1$,
we can regard $1/\beta=k_{\mbox{\scriptsize B}}T$ as the temperature.
Now, we pay attention to the fact that $T'_{0}(2)\simeq\pi/g$ for $l=2$ and it does not depend on $|\alpha|^{2}$ or $\theta(\beta)$.
This fact suggests that the two-photon JCM is insensitive to thermal effects.

The physical origin of the resilience observed only for the two-photon JCM is as follows.
Looking at Eq.~(\ref{matrix-elements-U-0}),
we notice that the operators $D$ and $D'$ cause phases of the wave function.
Because the explicit form of $D$ is given by Eq.~(\ref{explicit-forms-D-D-prime-operators}),
we obtain its eigenvalue
$D_{m}\simeq g^{2}m^{l}$ for $\Delta=0$ and $m\simeq|\alpha|^{2}\gg 1$.
Hence, the second line of Eq.~(\ref{Pe-approximation-form-0}) shows
that the phase of the cosine function is nearly equal to $2gm^{l/2}t$.
The peak of revival occurs at the time $T_{0}(l)$
when the phases produce constructive interference by satisfying the following relationship
\cite{Barnett1997}:
\begin{equation}
2gm^{l/2}T_{0}(l)-2g(m-1)^{l/2}T_{0}(l)=2\pi.
\end{equation}
When $l=2$, we can simplify this relationship as
\begin{equation}
2gmT_{0}(2)-2g(m-1)T_{0}(2)=2gT_{0}(2)=2\pi.
\end{equation}
Thus, we obtain $T_{0}(2)$ and $T'_{0}(2)$,
which are independent of $m\simeq|\alpha|^{2}$ and $\theta(\beta)$.
Therefore, the main reason for the particularity of the two-photon JCM is
the condition of the interference.

Now, we begin calculations of the low-temperature expansion of the probability to obtain the excited state of the atom.
Assuming that the temperature $1/\beta$ is constant during the time evolution,
we set the initial state
\begin{equation}
|\Psi_{\mbox{\scriptsize I}}(0)\rangle
=
|0(\Theta)\rangle_{\mbox{\scriptsize F}}
|\alpha;\theta\rangle_{\mbox{\scriptsize B}},
\end{equation}
\begin{eqnarray}
\cos\Theta(\beta)
&=&
[1+\exp(-\beta\omega_{0})]^{-1/2}, \nonumber \\
\sin\Theta(\beta)
&=&
\exp(-\beta\omega_{0}/2)[1+\exp(-\beta\omega_{0})]^{-1/2}, \nonumber \\
\cosh\theta(\beta)
&=&
[1-\exp(-\beta\omega)]^{-1/2}, \nonumber \\
\sinh\theta(\beta)
&=&
[\exp(\beta\omega)-1]^{-1/2},
\end{eqnarray}
where the subscripts F and B represent a fermion (the atom) and a boson (the cavity field).
The time evolution of the state is given by
\begin{eqnarray}
|\Psi_{\mbox{\scriptsize I}}(t)\rangle
&=&
\hat{U}(t)|\Psi_{\mbox{\scriptsize I}}(0)\rangle \nonumber \\
&=&
[U(t)\otimes\tilde{U}(t)]
|0(\Theta)\rangle_{\mbox{\scriptsize F}}
U_{\mbox{\scriptsize B}}(\theta)|\alpha\rangle_{\mbox{\scriptsize B}}
|\alpha^{*}\rangle_{\tilde{\mbox{\scriptsize B}}},
\end{eqnarray}
where
$|0(\Theta)\rangle_{\mbox{\scriptsize F}}$,
$|\alpha\rangle_{\mbox{\scriptsize B}}$, and
$|\alpha^{*}\rangle_{\tilde{\mbox{\scriptsize B}}}$
are defined in
${\cal H}_{\mbox{\scriptsize F}}
\otimes
\tilde{{\cal H}}_{\mbox{\scriptsize F}}$,
${\cal H}_{\mbox{\scriptsize B}}$,
and
$\tilde{{\cal H}}_{\mbox{\scriptsize B}}$,
respectively.
Representing orthogonal bases of fermions as four-component vectors,
\begin{eqnarray}
|0,\tilde{0}\rangle_{\mbox{\scriptsize F}}
&=&
(0,0,0,1)^{\mbox{\scriptsize T}}, \nonumber \\
|0,\tilde{1}\rangle_{\mbox{\scriptsize F}}
&=&
(0,0,1,0)^{\mbox{\scriptsize T}}, \nonumber \\
|1,\tilde{0}\rangle_{\mbox{\scriptsize F}}
&=&
(0,1,0,0)^{\mbox{\scriptsize T}}, \nonumber \\
|1,\tilde{1}\rangle_{\mbox{\scriptsize F}}
&=&
(1,0,0,0)^{\mbox{\scriptsize T}},
\end{eqnarray}
we write the thermal vacuum of the fermions as
\begin{eqnarray}
|0(\Theta)\rangle_{\mbox{\scriptsize F}}
&=&
\cos\Theta|0,\tilde{0}\rangle_{\mbox{\scriptsize F}}
+
\sin\Theta|1,\tilde{1}\rangle_{\mbox{\scriptsize F}} \nonumber \\
&=&
\left(
\begin{array}{c}
\sin\Theta \\
0 \\
0 \\
\cos\Theta
\end{array}
\right).
\end{eqnarray}
Its time evolution is written in the form
\begin{equation}
|\Psi_{\mbox{\scriptsize I}}(t)\rangle
=
[U(t)\otimes\tilde{U}(t)]
\left(
\begin{array}{c}
\sin\Theta U_{\mbox{\scriptsize B}}(\theta)
|\alpha\rangle_{\mbox{\scriptsize B}}|\alpha^{*}\rangle_{\tilde{\mbox{\scriptsize B}}} \\
0 \\
0 \\
\cos\Theta U_{\mbox{\scriptsize B}}(\theta)
|\alpha\rangle_{\mbox{\scriptsize B}}|\alpha^{*}\rangle_{\tilde{\mbox{\scriptsize B}}}
\label{time-evolution-interaction-picture-state-Psi-1}
\end{array}
\right).
\end{equation}
Then, the probability that we observe the excited state of the atom is given by
\begin{equation}
P_{\mbox{\scriptsize e}}(\Theta,\theta;t)
=
||_{\mbox{\scriptsize F}}\langle 1,\tilde{0}|\Psi_{\mbox{\scriptsize I}}(t)\rangle||^{2}
+
||_{\mbox{\scriptsize F}}\langle 1,\tilde{1}|\Psi_{\mbox{\scriptsize I}}(t)\rangle||^{2}.
\label{probability-excited-state-atom-original}
\end{equation}
In Appendix~\ref{section-perturbative-expansion-Pe}, we give the perturbative expansion of
$P_{\mbox{\scriptsize e}}(\Theta,\theta;t)$
up to the second-order term of $\theta(\beta)$. 

As explained above, we describe the time evolution of the multiphoton JCM
in contact with a heat bath using TFD.
This approach is equivalent to solving the Liouville-von Neumann equation,
\begin{equation}
\frac{\partial}{\partial t}\rho_{\mbox{\scriptsize I}}(t)
=
-i[\hat{H}_{\mbox{\scriptsize I}},\rho_{\mbox{\scriptsize I}}(t)],
\end{equation}
where
\begin{equation}
\rho_{\mbox{\scriptsize I}}(0)
=
\mbox{Tr}_{\tilde{\cal H}}[|\Psi_{\mbox{\scriptsize I}}(0)\rangle\langle\Psi_{\mbox{\scriptsize I}}(0)|],
\end{equation}
\begin{equation}
\hat{H}_{\mbox{\scriptsize I}}
=
g[c^{\dagger}a^{l}+c(a^{\dagger})^{l}]-\Delta c^{\dagger}c.
\end{equation}
Thus, the total system that we currently consider evolves in time at a constant temperature.
Hence, it is not affected by dissipation as an open system interacting
with an environment, and it develops in time reversibly.
Thus, the method discussed in the current paper cannot be
the dynamics of the Lindblad master equation or that of the Markovian master equation.
Under our TFD approach for the thermal equilibrium system,
the atom and the cavity field maintain their fermionic and bosonic thermal states,
with the Fermi-Dirac and the Bose-Einstein distributions, respectively, as reversible processes.
There are no differences between the TFD formalism and the Liouville-von Neumann equation.
However, if we apply the TFD approach to the problem of the multiphoton JCM
with a thermal coherent state,
we can derive the physical quantity as a perturbative series, that is,
powers of $\theta(\beta)$, with ease, as shown in Eq.~(\ref{probability-excited-state-atom-1}).
In contrast, if we want to obtain the perturbative series via the Liouville-von Neumann equation,
we must carry out a very complicated computation, and it is practically impossible.
This point is the reason why we choose the TFD formalism.

\section{\label{section-perturbative-calculation-relative-entropy}Perturbative calculations of thermal effects of the relative entropy of coherence}
The relative entropy of coherence of an arbitrary density matrix $\rho$ is given by
Eq.~(\ref{definition-relative-entropy-coherence-0}).
Writing elements of $\rho$ defined on ${\cal H}_{\mbox{\scriptsize F}}$ as
\begin{equation}
\rho
=
\left(
\begin{array}{cc}
\rho_{00} & \rho_{01} \\
\rho_{10} & \rho_{11} \\
\end{array}
\right),
\label{matrix-representation-rho}
\end{equation}
the relative entropy of coherence of the atom can be written in the form of Eq.~(\ref{definition-relative-entropy-coherence-0}) with
\begin{equation}
S(\rho_{\mbox{\scriptsize diag}})
=
-\rho_{00}\ln\rho_{00}-(1-\rho_{00})\ln(1-\rho_{00}),
\end{equation}
\begin{equation}
S(\rho)
=
-\lambda_{+}\ln\lambda_{+}-\lambda_{-}\ln\lambda_{-},
\end{equation}
\begin{equation}
\lambda_{\pm}
=
\frac{1}{2}[1\pm
\sqrt{1+4|\rho_{01}|^{2}-4\rho_{00}(1-\rho_{00})}].
\label{lambda-plus-minus}
\end{equation}
Because $\rho_{00}=P_{\mbox{\scriptsize e}}(\Theta,\theta;t)$ as defined
in Eq.~(\ref{probability-excited-state-atom-original})
and it is computed in Eqs.~(\ref{probability-excited-state-atom-1}),
(\ref{probability-excited-state-atom-2}), and (\ref{probability-excited-state-atom-3}),
the rest of what we must do is a calculation of $\rho_{01}$.
In Appendix~\ref{section-perturbative-expansion-rho-01},
we give the perturbative expansion of $\rho_{01}$ up to the second-order term of $\theta(\beta)$.

\section{\label{section-numerical-calculations}Numerical calculations}
\begin{figure}
\begin{center}
\includegraphics[width=\linewidth]{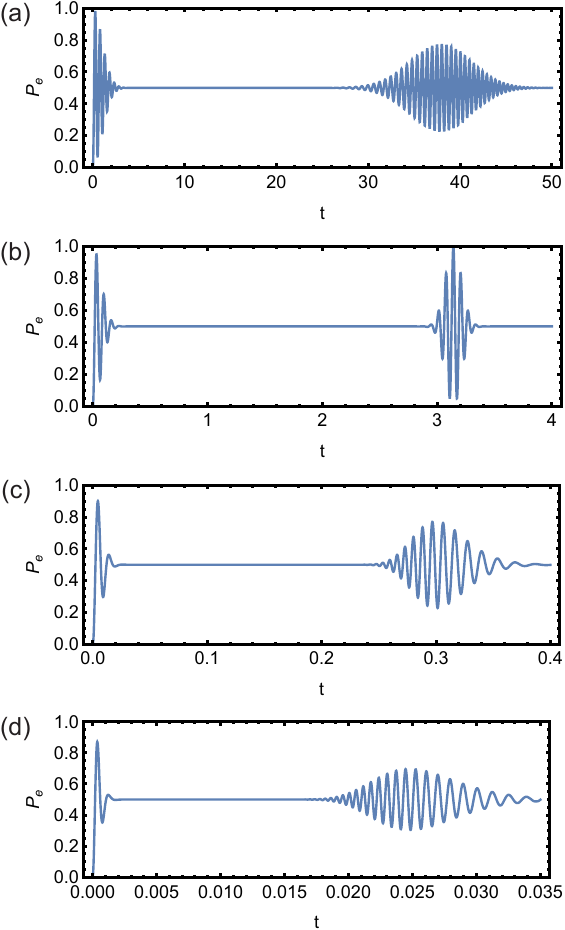}
\end{center}
\caption{Time variations of $P_{\mbox{\scriptsize e}}(\Theta,\theta;t)$
for $g=1$, $\omega_{0}=1$, $\omega=1$, and $1/\beta=0.1$.
(a)
A plot of $P_{\mbox{\scriptsize e}}(\Theta,\theta;t)$ as a function of $t$ with $l=1$, $\alpha=6$, and $\Delta=0$.
In this case, the zero-temperature approximate period is given by $T_{0}(1)=37.70$.
(b)
A plot of $P_{\mbox{\scriptsize e}}(\Theta,\theta;t)$ with $l=2$, $\alpha=7$, $\Delta=1$,
and $T_{0}(2)=3.142$.
(c)
A plot of $P_{\mbox{\scriptsize e}}(\Theta,\theta;t)$ with $l=3$, $\alpha=7$, $\Delta=2$,
and $T_{0}(3)=0.2992$.
(d)
A plot of $P_{\mbox{\scriptsize e}}(\Theta,\theta;t)$ with $l=4$, $\alpha=8$, $\Delta=3$,
and $T_{0}(4)=0.024{\,}54$.}
\label{figure01}
\end{figure}

Figure~\ref{figure01} shows time variations of $P_{\mbox{\scriptsize e}}(\Theta,\theta;t)$,
the probability of the excited state of the atom,
for the $l$-photon JCM with $l=1,2,3,4$
at temperature $1/\beta=0.1$.
In the curves in Figs.~\ref{figure01}(a)-\ref{figure01}(d),
we can observe revivals of the Rabi oscillations.

\begin{figure}
\begin{center}
\includegraphics[width=\linewidth]{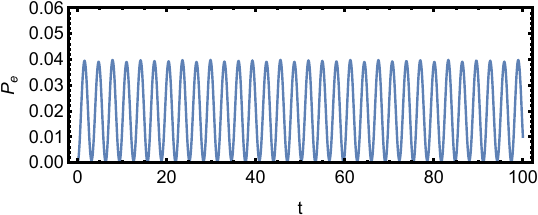}
\end{center}
\caption{
A plot of $P_{\mbox{\scriptsize e}}(\Theta,\theta;t)$
as a function of time $t$
for $l=1$, $g=1$, $\omega_{0}=1$, $\omega=1$, $1/\beta=0.1$, $\Delta=0$, and $\alpha=0.2$.
}
\label{figure02}
\end{figure}

In drawing Fig.~\ref{figure01}, we assume that the amplitude of the coherent state $\alpha$ is
large enough.
By contrast, if we set $|\alpha|\ll 1$, we cannot observe the collapse and revival
of the Rabi oscillations, as shown in Fig.~\ref{figure02}.
In Fig.~\ref{figure02}, we plot $P_{\mbox{\scriptsize e}}(\Theta,\theta;t)$ as a function of time for $l=1$ and $\alpha=0.2$.
Although $T_{0}(1)=1.257$, we cannot observe the collapse and revival of the Rabi oscillations in the range of $0\leq t\leq 100$.
Setting $g=1$, $\omega_{0}=1$, $\omega=1$, $1/\beta=0.1$, $0\leq t\leq 100$, and $\alpha=0.2$,
we obtain $P_{\mbox{\scriptsize e}}(\Theta,\theta;t)<8.0\times 10^{-4}$,
$P_{\mbox{\scriptsize e}}(\Theta,\theta;t)<5.0\times 10^{-5}$,
and
$P_{\mbox{\scriptsize e}}(\Theta,\theta;t)<5.0\times 10^{-5}$
for $l=2,3,4$, respectively.
We cannot find any signs of the collapses and revivals of the Rabi oscillations in those cases.

\begin{figure}
\begin{center}
\includegraphics[width=\linewidth]{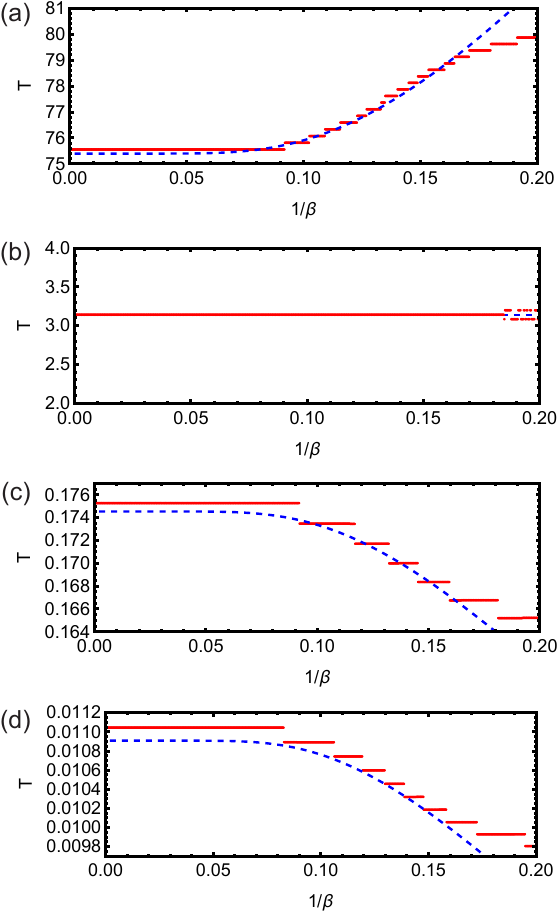}
\end{center}
\caption{Temperature dependence of the periods for $l=1,2,3,4$ with $\alpha=12$, $g=1$, $\omega_{0}=1$, and $\omega=1$.
Red lines represent the periods derived from time variations of $P_{\mbox{\scriptsize e}}(\Theta,\theta;t)$.
Dashed blue curves represent $T'_{0}(l)$ defined by Eq.~(\ref{Period-beta-dependence-approximate-1}).
(a)
$l=1$, and $\Delta=0$.
(b)
$l=2$, and $\Delta=1$.
(c)
$l=3$, and $\Delta=2$.
(d)
$l=4$, and $\Delta=3$.
In (b), red lines become unstable for $0.18\leq 1/\beta$
because the second-order perturbation theory is not effective for this range of $1/\beta$.}
\label{figure03}
\end{figure}

Figure~\ref{figure03} shows the temperature dependence of the period
of the collapse and revival of the Rabi oscillations for the $l$-photon JCM with $l=1,2,3,4$.
Red lines represent the periods derived from time variations of $P_{\mbox{\scriptsize e}}(\Theta,\theta;t)$.
Dashed blue curves represent $T'_{0}(l)$ obtained by intuitive approximation
in Eq.~(\ref{Period-beta-dependence-approximate-1}).

In Figs.~\ref{figure03}(a), \ref{figure03}(c), and \ref{figure03}(d),
red curves show discrete values of the periods.
The reason for this phenomenon is as follows.
For example, in Fig.~\ref{figure03}(a),
the period of the Rabi oscillation defined in Eq.~(\ref{period-Rabi-oscillations-l-1}) is given by
$\tau_{1}=\pi/(g|\alpha|)=0.2618$.
Thus, differences in the discrete periods of the collapse and revival of the Rabi oscillations are multiples of $\tau_{1}$.
Similarly, we observe discrete periods in Figs.~\ref{figure03}(c) and \ref{figure03}(d).
By contrast, in Fig.~\ref{figure03}(b),
the red line and the dashed blue curve match each other at the constant value $T=3.142$ for any temperature.
This fact shows that the two-photon JCM is resistant to thermal noise.

Looking at Figs.~\ref{figure03}(a)-\ref{figure03}(d),
we note that the red lines and the dashed blue curves coincide well
in $0\leq 1/\beta\leq 0.16$.
We can regard these values of $1/\beta$ as an effective range where the second-order perturbation theory is valid.

\begin{figure}
\begin{center}
\includegraphics[width=0.92\linewidth]{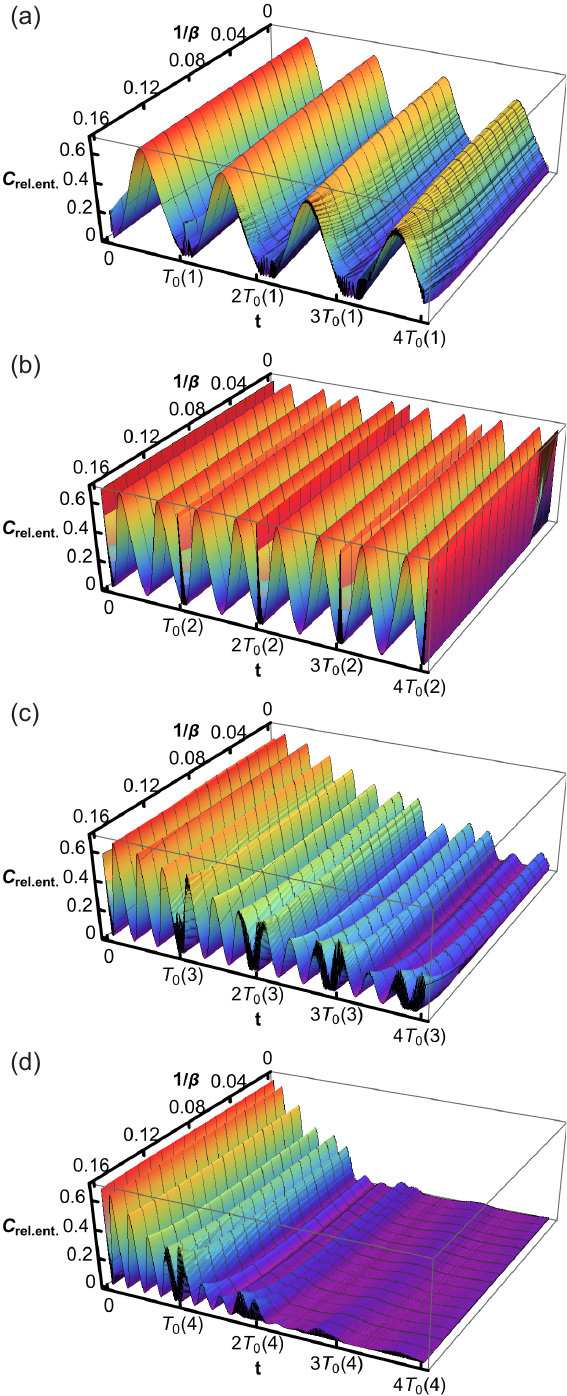}
\end{center}
\caption{Three-dimensional plots of the relative entropy of coherence $C_{\mbox{\scriptsize rel.ent.}}$
as a function of the time $t$ and the temperature $1/\beta$, with $g=1$, $\omega_{0}=1$, $\omega=1$, and $\alpha=12$.
We set parameters
(a) $l=1$ and $\Delta=0$,
(b) $l=2$ and $\Delta=1$,
(c) $l=3$ and $\Delta=2$,
and
(d) $l=4$ and $\Delta=3$.
Ticks on the axes for $1/\beta$ in the graphs increase from the back to the front.
For ticks on the axes for $t$ in the graphs, we put $T_{0}(l)$ defined in Eq.~(\ref{period-collapses-revivals-Rabi-oscillations-0}) for $l=1,2,3,4$.
Looking at (c) and (d),
we note that distinct thermal fluctuations arise around $t\simeq T_{0}(l),2T_{0}(l),3T_{0}(l)$ for $l=3,4$ and $\beta\simeq 0.16$.}
\label{figure04}
\end{figure}

Figures.~\ref{figure04}(a)-\ref{figure04}(d) represent three-dimensional plots of the relative entropy of coherence $C_{\mbox{\scriptsize rel.ent.}}$
as a function of time $t$ and temperature $1/\beta$ for $l=1,2,3,4$, respectively.
To have the second-order perturbation theory be effective,
we use the range $0\leq\beta\leq 0.16$ obtained in Fig.~\ref{figure03}.
Looking at Figs.~\ref{figure04}(a), \ref{figure04}(c), and \ref{figure04}(d), we note that $C_{\mbox{\scriptsize rel.ent.}}$ decays as time $t$ proceeds even at zero temperature ($1/\beta=0$).
Moreover, in these plots, fluctuations of $C_{\mbox{\scriptsize rel.ent.}}$ increase as temperature $1/\beta$ becomes larger.
By contrast, in Fig.~\ref{figure04}(b) for $l=2$,
$C_{\mbox{\scriptsize rel.ent.}}$ does not decay or suffer from thermal noise
for arbitrary temperature.
Thus, we can conclude that $C_{\mbox{\scriptsize rel.ent.}}$ in the two-photon JCM is stable under thermal noise.

In Fig.~\ref{figure04}, we set $\alpha=12$ to calculate $C_{\mbox{\scriptsize rel.ent.}}$
with $l=1,2,3,4$.
In contrast, if we let $\alpha$ be a small value, we cannot obtain any interesting results.
For $l=1$, $g=1$, $\omega_{0}=1$, $\omega=1$, $\alpha=0.2$, $0\leq 1/\beta\leq 0.16$,
and $0\leq t\leq 100T_{0}(1)$,
numerical calculations show that $C_{\mbox{\scriptsize rel.ent.}}$ is less than $0.2$,
oscillates in time with a period $\pi$,
and does not suffer from thermal effects.
With the same physical parameters, we obtain
$C_{\mbox{\scriptsize rel.ent.}}<6.0\times 10^{-3}$,
$C_{\mbox{\scriptsize rel.ent.}}<1.2\times 10^{-4}$,
and
$C_{\mbox{\scriptsize rel.ent.}}<1.8\times 10^{-6}$
for $l=2,3,4$, respectively.

In this section, we numerically compute
$P_{\mbox{\scriptsize e}}(\Theta,\theta;t)$, the period, and $C_{\mbox{\scriptsize rel.ent.}}$
while setting $g=\omega=\omega_{0}=1.0$,
which is a large coupling strength.
The reason why we choose such a strong-coupling regime is that we want to let $T_{0}(2)$ be equal to $\pi$.
However, for example, if we set $g=0.1$, that is to say,
a weak coupling strength, we also obtain a similar
$P_{\mbox{\scriptsize e}}(\Theta,\theta;t)$, period, and $C_{\mbox{\scriptsize rel.ent.}}$,
as shown in Figs.~\ref{figure01}, \ref{figure03}, and \ref{figure04},
and their behavior is not modified significantly;
thus, the physics of the system does not change essentially.
If we change the coupling constant $g$ as $g \to g'=cg$ with a constant $c$,
we can absorb it by redefining the time variable $t$ as $t'=t/c$.
This alteration of the scale of the time variable is effective for the interaction picture
given by Eqs.~(\ref{definition-U-t-0}) and (\ref{definition-U-hat-t-0}).
Thus, the change in the coupling strength $g$ does not cause a modification of the physics of the system.

For this section, we performed the numerical calculations with C++ programs.
We evaluated
$P_{\mbox{\scriptsize e}}(\Theta,\theta;t)$, the periods, and $C_{\mbox{\scriptsize rel.ent.}}$
according to Eqs.~(\ref{definition-relative-entropy-coherence-0}),
(\ref{matrix-representation-rho})-(\ref{lambda-plus-minus}),
(\ref{probability-excited-state-atom-1})-(\ref{S-1-2-k-t}),
and (\ref{rho-01-1})-(\ref{definition-u-n}).
Obtaining these values numerically,
we must compute the series,
shown in Eq.~(\ref{S-1-2-k-t}), for example.
To calculate those series, we carry out summations from $n=0$ to $n=n_{\mbox{\scriptsize Max}}$ instead of $n\to+\infty$, and we set $n_{\mbox{\scriptsize Max}}=110$ for Fig.~\ref{figure01},
$n_{\mbox{\scriptsize Max}}=80$ for Fig.~\ref{figure02},
and $n_{\mbox{\scriptsize Max}}=250$ for Figs.~\ref{figure03} and \ref{figure04}.
We do not perform anything special other than what was discussed above for the numerical calculations in this section.

\section{\label{section-discussion}Discussion}
In this paper, we investigated the thermal effects of the period of the collapse and revival of the Rabi oscillations and the relative entropy of coherence for a multiphoton JCM
whose initial state of the cavity field is given by coherent light.
We showed that these physical quantities hardly suffer from thermal noise using numerical calculations for the two-photon JCM.
This insensitivity of the two-photon JCM to thermal effects will have a wide range of applications in quantum information processes.
We can expect that quantum devices implemented with the two-photon JCM are resistant to thermal noise.
In Ref.~\cite{Kuhn1999}, an on-demand single-photon source implemented with a strongly coupled atom-cavity system was proposed.
The functions of this device are based on stimulated Raman adiabatic passage (STIRAP) \cite{Vitanov2017},
and the single-photon JCM plays an important role in realizing STIRAP.
Thus, for example, we may construct an on-demand photon-pair source with the two-photon JCM.
For analyses of thermal fluctuations induced in the multiphoton JCM,
we use TFD as a powerful and convenient tool.
Hence, our work is an important application of TFD.
Because experimental realization of the multiphoton JCM has become a real possibility recently,
our results will contribute to the development of devices for quantum information processing.

\section*{Acknowledgment}
This work was supported by MEXT Quantum Leap Flagship Program Grant No. JPMXS0120351339.

\appendix

\section{\label{section-derivation-equation-15}The derivation of
Eq.~(\ref{approximation-derivation-Pe})}
Using Eq.~(\ref{approx-m_to-l-over-2}), we can obtain
Eq.~(\ref{approximation-derivation-Pe}) as follows:
\begin{eqnarray}
&&
\sum_{m=0}^{\infty}
\frac{|\alpha|^{2m}}{m!}\cos(2gm^{l/2}t) \nonumber \\
&\simeq &
\frac{1}{2}\exp(ig|\alpha|^{l}t)
\sum_{m=0}^{\infty}\frac{1}{m!}[|\alpha|^{2}\exp(ig|\alpha|^{l-2}lt)]^{m}  \nonumber \\
&&
+
\frac{1}{2}\exp(-ig|\alpha|^{l}t)
\sum_{m=0}^{\infty}\frac{1}{m!}[|\alpha|^{2}\exp(-ig|\alpha|^{l-2}lt)]^{m}  \nonumber \\
&=&
\frac{1}{2}\exp(ig|\alpha|^{l}t)
\exp[|\alpha|^{2}\exp(ig|\alpha|^{l-2}lt)] \nonumber \\
&&
+
\frac{1}{2}\exp(-ig|\alpha|^{l}t)
\exp[|\alpha|^{2}\exp(-ig|\alpha|^{l-2}lt)] \nonumber \\
&=&
\exp[|\alpha|^{2}\cos(g|\alpha|^{l-2}lt)] \nonumber \\
&&
\times
\{\cos(g|\alpha|^{l}t)\cos[|\alpha|^{2}\sin(g|\alpha|^{l-2}lt)] \nonumber \\
&&
-
\sin(g|\alpha|^{l}t)\sin[|\alpha|^{2}\sin(g|\alpha|^{l-2}lt)]\} \nonumber \\
&=&
\exp[|\alpha|^{2}\cos(g|\alpha|^{l-2}lt)] \nonumber \\
&&
\times
\cos[g|\alpha|^{l}t+|\alpha|^{2}\sin(g|\alpha|^{l-2}lt)].
\end{eqnarray}

\section{\label{section-review-TFD}A brief review of TFD}
In TFD, we prepare twin Hilbert spaces ${\cal H}\otimes\tilde{{\cal H}}$.
For boson systems, the orthogonal bases of ${\cal H}_{\mbox{\scriptsize B}}$ and $\tilde{{\cal H}}_{\mbox{\scriptsize B}}$
are given by
$\{|n\rangle_{\mbox{\scriptsize B}}: n=0,1,2,...\}$
and
$\{|\tilde{n}\rangle_{\tilde{\mbox{\scriptsize B}}}: \tilde{n}=0,1,2,...\}$, respectively.
Annihilation operators of the bosons on ${\cal H}_{\mbox{\scriptsize B}}$ and $\tilde{{\cal H}}_{\mbox{\scriptsize B}}$ are given
by $a$ and $\tilde{a}$, respectively.
They satisfy the commutation relations
$[a,a^{\dagger}]=[\tilde{a},\tilde{a}^{\dagger}]=1$
and
$[a,\tilde{a}]=[a,\tilde{a}^{\dagger}]=0$.
Here, we introduce the inverse of the temperature $\beta=1/(k_{\mbox{\scriptsize B}}T)$,
where $k_{\mbox{\scriptsize B}}$ denotes the Boltzmann constant.
We define the Bogoliubov transformation for the bosons as
\begin{equation}
U_{\mbox{\scriptsize B}}(\theta)=\exp[i\theta(\beta)G_{\mbox{\scriptsize B}}],
\end{equation}
\begin{equation}
G_{\mbox{\scriptsize B}}=i(a\tilde{a}-\tilde{a}^{\dagger}a^{\dagger}),
\end{equation}
\begin{eqnarray}
\cosh\theta(\beta)&=&[1-\exp(-\beta\epsilon)]^{-1/2}, \nonumber \\
\sinh\theta(\beta)&=&[\exp(\beta\epsilon)-1]^{-1/2},
\label{definition-theta-beta}
\end{eqnarray}
where
$\epsilon=\omega$ and $\omega$ represents the angular frequency of the boson.
Applying the Bogoliubov transformation to $a$ and $\tilde{a}$, we obtain
\begin{eqnarray}
a
&\to&
a(\theta)=U_{\mbox{\scriptsize B}}(\theta)aU^{\dagger}_{\mbox{\scriptsize B}}(\theta) \nonumber \\
&&
\quad\quad
=\cosh\theta(\beta)a-\sinh\theta(\beta)\tilde{a}^{\dagger}, \nonumber \\
\tilde{a}
&\to&
\tilde{a}(\theta)=U_{\mbox{\scriptsize B}}(\theta)\tilde{a}U^{\dagger}_{\mbox{\scriptsize B}}(\theta) \nonumber \\
&&
\quad\quad
=\cosh\theta(\beta)\tilde{a}-\sinh\theta(\beta)a^{\dagger},
\label{definition-a-theta}
\end{eqnarray}
and they satisfy the commutation relations
\begin{eqnarray}
&&
[a(\theta),a^{\dagger}(\theta)]=[\tilde{a}(\theta),\tilde{a}^{\dagger}(\theta)]=1, \nonumber \\
&&
[a(\theta),\tilde{a}(\theta)]=[a(\theta),\tilde{a}^{\dagger}(\theta)]=0.
\end{eqnarray}

Defining the zero-temperature vacuum state
$|0,\tilde{0}\rangle_{\mbox{\scriptsize B}}
=|0\rangle_{\mbox{\scriptsize B}}\otimes|\tilde{0}\rangle_{\tilde{\mbox{\scriptsize B}}}
\in{\cal H}_{\mbox{\scriptsize B}}\otimes\tilde{{\cal H}}_{\mbox{\scriptsize B}}$,
we can derive the finite-temperature vacuum state as
\begin{eqnarray}
|0(\theta)\rangle_{\mbox{\scriptsize B}}
&=&
U_{\mbox{\scriptsize B}}(\theta)|0,\tilde{0}\rangle_{\mbox{\scriptsize B}} \nonumber \\
&=&
\exp(-\ln\cosh\theta)\exp[(\tanh\theta)a^{\dagger}\tilde{a}^{\dagger}] \nonumber \\
&&
\times
|0,\tilde{0}\rangle_{\mbox{\scriptsize B}},
\end{eqnarray}
\begin{equation}
a(\theta)|0(\theta)\rangle_{\mbox{\scriptsize B}}
=
\tilde{a}(\theta)|0(\theta)\rangle_{\mbox{\scriptsize B}}
=
0.
\end{equation}
The physical meaning of $|0(\theta)\rangle_{\mbox{\scriptsize B}}$ is as follows.
Tracing out the degrees of freedom of $\tilde{{\cal H}}_{\mbox{\scriptsize B}}$,
we obtain the density matrix of ${\cal H}_{\mbox{\scriptsize B}}$ as
\begin{eqnarray}
\rho_{\mbox{\scriptsize B}}(\theta)
&=&
\mbox{Tr}_{\tilde{\cal H}_{\mbox{\tiny B}}}|0(\theta)\rangle_{\mbox{\scriptsize B}}{}_{\mbox{\scriptsize B}}\langle 0(\theta)| \nonumber \\
&=&
(1-e^{-\beta\epsilon})\sum_{n=0}^{\infty}e^{-n\beta\epsilon}|n\rangle_{\mbox{\scriptsize B}}{}_{\mbox{\scriptsize B}}\langle n|,
\end{eqnarray}
and we can regard it as the Bose-Einstein distribution.
Hence, introducing the second Hilbert space $\tilde{\cal H}_{\mbox{\scriptsize B}}$ and considering the two-mode squeezed vacuum state
$|0(\theta)\rangle_{\mbox{\scriptsize B}}$,
the classical statistical theory is naturally induced.

Next, we discuss the finite-temperature fermion system.
We consider twin Hilbert spaces ${\cal H}_{\mbox{\scriptsize F}}$ and $\tilde{\cal H}_{\mbox{\scriptsize F}}$,
whose orthogonal bases are given by
$\{|0\rangle_{\mbox{\scriptsize F}},|1\rangle_{\mbox{\scriptsize F}}\}$
and
$\{|\tilde{0}\rangle_{\tilde{\mbox{\scriptsize F}}},|\tilde{1}\rangle_{\tilde{\mbox{\scriptsize F}}}\}$, respectively.
We define annihilation operators of ${\cal H}_{\mbox{\scriptsize F}}$ and $\tilde{\cal H}_{\mbox{\scriptsize F}}$
as $c$ and $\tilde{c}$, respectively.
They satisfy the commutation relations
$\{c,c^{\dagger}\}=\{\tilde{c},\tilde{c}^{\dagger}\}=1$
and
$\{c,\tilde{c}\}=\{c,\tilde{c}^{\dagger}\}=0$.
The Bogoliubov transformation in ${\cal H}_{\mbox{\scriptsize F}}$ and $\tilde{\cal H}_{\mbox{\scriptsize F}}$ is written in the form
\begin{equation}
U_{\mbox{\scriptsize F}}(\theta)=\exp[i\theta(\beta)G_{\mbox{\scriptsize F}}],
\end{equation}
\begin{equation}
G_{\mbox{\scriptsize F}}=i(c\tilde{c}-\tilde{c}^{\dagger}c^{\dagger}),
\end{equation}
\begin{eqnarray}
\cos\theta(\beta)&=&[1+\exp(-\beta\epsilon)]^{-1/2}, \nonumber \\
\sin\theta(\beta)&=&\exp(-\beta\epsilon/2)[1+\exp(-\beta\epsilon)]^{-1/2},
\end{eqnarray}
where $\epsilon=\omega$ and $\omega$ is an angular frequency of the fermions.
Applying the Bogoliubov transformation to $c$ and $\tilde{c}$,
we obtain
\begin{eqnarray}
c
&\to&
c(\theta)=U_{\mbox{\scriptsize F}}(\theta)cU^{\dagger}_{\mbox{\scriptsize F}}(\theta) \nonumber \\
&&
\quad\quad
=\cos\theta(\beta)c+\sin\theta(\beta)\tilde{c}^{\dagger}, \nonumber \\
\tilde{c}
&\to&
\tilde{c}(\theta)=U_{\mbox{\scriptsize F}}(\theta)\tilde{c}U^{\dagger}_{\mbox{\scriptsize F}}(\theta) \nonumber \\
&&
\quad\quad
=\cos\theta(\beta)\tilde{c}-\sin\theta(\beta)c^{\dagger},
\end{eqnarray}
and their commutation relations are given by
\begin{eqnarray}
&&
\{c(\theta),c^{\dagger}(\theta)\}=\{\tilde{c}(\theta),\tilde{c}^{\dagger}(\theta)\}=1, \nonumber \\
&&
\{c(\theta),\tilde{c}(\theta)\}=\{c(\theta),\tilde{c}^{\dagger}(\theta)\}=0.
\end{eqnarray}

We define the zero-temperature vacuum state of
${\cal H}_{\mbox{\scriptsize F}}\otimes\tilde{{\cal H}}_{\mbox{\scriptsize F}}$
as
$|0,\tilde{0}\rangle_{\mbox{\scriptsize F}}$.
Then, the finite-temperature vacuum state is written as
\begin{eqnarray}
|0(\theta)\rangle_{\mbox{\scriptsize F}}
&=&
U_{\mbox{\scriptsize F}}(\theta)|0,\tilde{0}\rangle_{\mbox{\scriptsize F}} \nonumber \\
&=&
[\cos\theta+(\sin\theta)c^{\dagger}\tilde{c}^{\dagger}]|0,\tilde{0}\rangle_{\mbox{\scriptsize F}},
\end{eqnarray}
\begin{equation}
c(\theta)|0(\theta)\rangle_{\mbox{\scriptsize F}}
=
\tilde{c}(\theta)|0(\theta)\rangle_{\mbox{\scriptsize F}}
=
0.
\end{equation}
Tracing out the degrees of freedom of $\tilde{{\cal H}}_{\mbox{\scriptsize F}}$,
we obtain the density matrix of ${\cal H}_{\mbox{\scriptsize F}}$ as
\begin{eqnarray}
\rho_{\mbox{\scriptsize F}}(\theta)
&=&
\mbox{Tr}_{\tilde{\cal H}_{\mbox{\tiny F}}}|0(\theta)\rangle_{\mbox{\scriptsize F}}{}_{\mbox{\scriptsize F}}\langle 0(\theta)| \nonumber \\
&=&
(1+e^{-\beta\epsilon})^{-1}|0\rangle_{\mbox{\scriptsize F}}{}_{\mbox{\scriptsize F}}\langle 0| \nonumber \\
&&
+
e^{-\beta\epsilon}
(1+e^{-\beta\epsilon})^{-1}|1\rangle_{\mbox{\scriptsize F}}{}_{\mbox{\scriptsize F}}\langle 1|,
\end{eqnarray}
and we can regard it as the Fermi-Dirac distribution.

Because we introduce the fictional Hilbert space $\tilde{{\cal H}}$ in addition to the real Hilbert space ${\cal H}$,
there are twice as many degrees of freedom of the system.
Thus, to describe a genuine system, we need to apply restrictions to states.
Hence, we make states invariant under the tilde conjugation,
\begin{eqnarray}
(XY)\tilde{\;\;}
&=&
\tilde{X}\tilde{Y}, \nonumber \\
(\xi_{1}X+\xi_{2}Y)\tilde{\;\;}
&=&
\xi_{1}^{*}\tilde{X}+\xi_{2}^{*}\tilde{Y}, \nonumber \\
(X^{\dagger})\tilde{\;\;}
&=&
\tilde{X}^{\dagger}, \nonumber \\
(\tilde{X})\tilde{\;\;}
&=&
\sigma X,
\end{eqnarray}
\begin{equation}
\sigma
=
\left\{
\begin{array}{ll}
1 & \mbox{(boson),}\\
-1 & \mbox{(fermion),}
\end{array}
\right.
\end{equation}
where $X$ and $Y$ are arbitrary operators defined on ${\cal H}_{\mbox{\scriptsize B}}$ and/or ${\cal H}_{\mbox{\scriptsize F}}$
and $\xi_{1}$ and $\xi_{2}$ are arbitrary complex numbers.
Because we want to study the thermal effects of the collapses and the revivals of the multiphoton JCM,
we define a thermal coherent state in the form
\cite{Mann1989,Kireev1989}
\begin{equation}
|\alpha;\theta\rangle_{\mbox{\scriptsize B}}
=
U_{\mbox{\scriptsize B}}(\theta)|\alpha\rangle_{\mbox{\scriptsize B}}|\alpha^{*}\rangle_{\tilde{\mbox{\scriptsize B}}},
\label{definition-thermal-coherent-state-1}
\end{equation}
where $|\alpha\rangle_{\mbox{\scriptsize B}}$ and $|\alpha^{*}\rangle_{\tilde{\mbox{\scriptsize B}}}$
are coherent states defined in
${\cal H}_{\mbox{\scriptsize B}}$ and $\tilde{\cal H}_{\mbox{\scriptsize B}}$, respectively.
This thermal coherent state is invariant under the tilde conjugation.

\section{\label{section-perturbative-expansion-Pe}The perturbative expansion
of $P_{\mbox{\scriptsize e}}(\Theta,\theta;t)$}
The unitary operators $U(t)$ and $\tilde{U}(t)$ are given by $2\times 2$ matrices,
\begin{equation}
U(t)
=
\left(
\begin{array}{cc}
u_{00} & u_{01} \\
u_{10} & u_{11}
\end{array}
\right),
\quad
\tilde{U}(t)
=
\left(
\begin{array}{cc}
\tilde{u}_{00} & \tilde{u}_{01} \\
\tilde{u}_{10} & \tilde{u}_{11}
\end{array}
\right).
\label{unitary-U-tilde-U-definition-0}
\end{equation}
Substituting Eq.~(\ref{unitary-U-tilde-U-definition-0})
into Eq.~(\ref{time-evolution-interaction-picture-state-Psi-1}),
we obtain
\begin{equation}
|\Psi_{\mbox{\scriptsize I}}(t)\rangle
=
(
\psi_{0\tilde{0}},
\psi_{0\tilde{1}},
\psi_{1\tilde{0}},
\psi_{1\tilde{1}}
)^{\mbox{\scriptsize T}},
\end{equation}
\begin{eqnarray}
\psi_{0\tilde{0}}
&=&
(\sin\Theta u_{00}\tilde{u}_{00}
+\cos\Theta u_{01}\tilde{u}_{01})
U_{\mbox{\scriptsize B}}(\theta)
|\alpha\rangle{\mbox{\scriptsize B}}
|\alpha^{*}\rangle{\tilde{\mbox{\scriptsize B}}}, \nonumber \\
\psi_{0\tilde{1}}
&=&
(\sin\Theta u_{00}\tilde{u}_{10}
+\cos\Theta u_{01}\tilde{u}_{11})
U_{\mbox{\scriptsize B}}(\theta)
|\alpha\rangle{\mbox{\scriptsize B}}
|\alpha^{*}\rangle{\tilde{\mbox{\scriptsize B}}}, \nonumber \\
\psi_{1\tilde{0}}
&=&
(\sin\Theta u_{10}\tilde{u}_{00}
+\cos\Theta u_{11}\tilde{u}_{01})
U_{\mbox{\scriptsize B}}(\theta)
|\alpha\rangle{\mbox{\scriptsize B}}
|\alpha^{*}\rangle{\tilde{\mbox{\scriptsize B}}}, \nonumber \\
\psi_{1\tilde{1}}
&=&
(\sin\Theta u_{10}\tilde{u}_{10}
+\cos\Theta u_{11}\tilde{u}_{11})
U_{\mbox{\scriptsize B}}(\theta)
|\alpha\rangle{\mbox{\scriptsize B}}
|\alpha^{*}\rangle{\tilde{\mbox{\scriptsize B}}}. \nonumber \\
\end{eqnarray}
The probability is given by
\begin{eqnarray}
P_{\mbox{\scriptsize e}}(\Theta,\theta;t)
&=&
||\psi_{0\tilde{0}}||^{2}
+
||\psi_{0\tilde{1}}||^{2} \nonumber \\
&=&
{}_{\mbox{\scriptsize B}}\langle\alpha|_{\tilde{\mbox{\scriptsize B}}}\langle\alpha^{*}|
U_{\mbox{\scriptsize B}}^{\dagger}(\theta)
(\sin^{2}\Theta u^{\dagger}_{00}u_{00} \nonumber \\
&&
+\cos^{2}\Theta u^{\dagger}_{01}u_{01})
U_{\mbox{\scriptsize B}}(\theta)
|\alpha\rangle_{\mbox{\scriptsize B}}|\alpha^{*}\rangle_{\tilde{\mbox{\scriptsize B}}},
\label{probability-excited-state-atom-0}
\end{eqnarray}
where we use
\begin{eqnarray}
\tilde{u}^{\dagger}_{00}\tilde{u}_{00}+\tilde{u}^{\dagger}_{10}\tilde{u}_{10}
&=&
\tilde{u}^{\dagger}_{01}\tilde{u}_{01}+\tilde{u}^{\dagger}_{11}\tilde{u}_{11}
=1, \nonumber \\
\tilde{u}^{\dagger}_{00}\tilde{u}_{01}+\tilde{u}^{\dagger}_{10}\tilde{u}_{11}
&=&
\tilde{u}^{\dagger}_{01}\tilde{u}_{00}+\tilde{u}^{\dagger}_{11}\tilde{u}_{10}
=0.
\end{eqnarray}
Because $U_{\mbox{\scriptsize B}}(\theta)=\exp[-\theta(a\tilde{a}-\tilde{a}^{\dagger}a^{\dagger})]$,
using the Baker-Campbell-Hausdorff formula
\begin{equation}
e^{X}Ye^{-X}=Y+[X,Y]+(1/2)[X,[X,Y]]+...,
\end{equation}
we obtain
\begin{equation}
P_{\mbox{\scriptsize e}}(\Theta,\theta;t)
=
\sum_{n=0}^{\infty}\frac{\theta(\beta)^{n}}{n!}
P_{\mbox{\scriptsize e}}^{(n)}(\Theta;t),
\label{probability-excited-state-atom-1}
\end{equation}
\begin{equation}
P_{\mbox{\scriptsize e}}^{(n)}(\Theta;t)
=
\sin^{2}\Theta P_{\mbox{\scriptsize e},1}^{(n)}(t)
+
\cos^{2}\Theta P_{\mbox{\scriptsize e},2}^{(n)}(t),
\label{probability-excited-state-atom-2}
\end{equation}
\begin{eqnarray}
P_{\mbox{\scriptsize e},1}^{(0)}(t)
&=&
{\cal P}^{(0)}(u^{\dagger}_{00}u_{00}),
\quad
P_{\mbox{\scriptsize e},2}^{(0)}(t)
=
{\cal P}^{(0)}(u^{\dagger}_{01}u_{01}), \nonumber \\
P_{\mbox{\scriptsize e},1}^{(1)}(t)
&=&
{\cal P}^{(1)}(u^{\dagger}_{00}u_{00}),
\quad
P_{\mbox{\scriptsize e},2}^{(1)}(t)
=
{\cal P}^{(1)}(u^{\dagger}_{01}u_{01}), \nonumber \\
P_{\mbox{\scriptsize e},1}^{(2)}(t)
&=&
{\cal P}^{(2)}(u^{\dagger}_{00}u_{00}),
\quad
P_{\mbox{\scriptsize e},2}^{(2)}(t)
=
{\cal P}^{(2)}(u^{\dagger}_{01}u_{01}), \nonumber \\
\label{probability-excited-state-atom-3}
\end{eqnarray}
where
\begin{eqnarray}
{\cal P}^{(0)}(X)
&=&
{}_{\mbox{\scriptsize B}}\langle\alpha|_{\tilde{\mbox{\scriptsize B}}}\langle\alpha^{*}|
X
|\alpha\rangle_{\mbox{\scriptsize B}}|\alpha^{*}\rangle_{\tilde{\mbox{\scriptsize B}}}, \nonumber \\
{\cal P}^{(1)}(X)
&=&
{}_{\mbox{\scriptsize B}}\langle\alpha|_{\tilde{\mbox{\scriptsize B}}}\langle\alpha^{*}|
[
a\tilde{a}-\tilde{a}^{\dagger}a^{\dagger},
X
]
|\alpha\rangle_{\mbox{\scriptsize B}}|\alpha^{*}\rangle_{\tilde{\mbox{\scriptsize B}}}, \nonumber \\
{\cal P}^{(2)}(X)
&=&
{}_{\mbox{\scriptsize B}}\langle\alpha|_{\tilde{\mbox{\scriptsize B}}}\langle\alpha^{*}|
[
a\tilde{a}-\tilde{a}^{\dagger}a^{\dagger},
[
a\tilde{a}-\tilde{a}^{\dagger}a^{\dagger},
X
]
] \nonumber \\
&& \times
|\alpha\rangle_{\mbox{\scriptsize B}}|\alpha^{*}\rangle_{\tilde{\mbox{\scriptsize B}}}
\label{probability-excited-state-atom-4}
\end{eqnarray}
for an arbitrary operator $X$.
Finally, we obtain
\begin{eqnarray}
P_{\mbox{\scriptsize e},1}^{(0)}(t)
&=&
e^{-|\alpha|^{2}}S_{1}(0), \nonumber \\
P_{\mbox{\scriptsize e},2}^{(0)}(t)
&=&
g^{2}|\alpha|^{2l}e^{-|\alpha|^{2}}S_{2}(0), \nonumber \\
P_{\mbox{\scriptsize e},1}^{(1)}(t)
&=&
-2|\alpha|^{2}e^{-|\alpha|^{2}}[S_{1}(0)-S_{1}(1)], \nonumber \\
P_{\mbox{\scriptsize e},2}^{(1)}(t)
&=&
2g^{2}|\alpha|^{2l}e^{-|\alpha|^{2}}
[(l-|\alpha|^{2})S_{2}(0)+|\alpha|^{2}S_{2}(1)], \nonumber \\
P_{\mbox{\scriptsize e},1}^{(2)}(t)
&=&
2e^{-|\alpha|^{2}}
[
-(1+|\alpha|^{2}-2|\alpha|^{4})S_{1}(0) \nonumber \\
&&
+(1-4|\alpha|^{4})S_{1}(1)
+|\alpha|^{2}(1+2|\alpha|^{2})S_{1}(2)], \nonumber \\
P_{\mbox{\scriptsize e},2}^{(2)}(t)
&=&
2g^{2}e^{-|\alpha|^{2}}(1+2|\alpha|^{2})|\alpha|^{2(l-1)} \nonumber \\
&& \times
\{
[l^{2}-(1+2l)|\alpha|^{2}+|\alpha|^{4}]S_{2}(0) \nonumber \\
&&
-
|\alpha|^{2}(-1-2l+2|\alpha|^{2})S_{2}(1) \nonumber \\
&&
+
|\alpha|^{4}S_{2}(2)\},
\end{eqnarray}
where
\begin{eqnarray}
S_{1}(k)
&=&
\sum_{n=0}^{\infty}
\frac{|\alpha|^{2n}}{n!}
\Big[
\cos^{2}(\sqrt{D_{n+k}}t) \nonumber \\
&&
+
\Big(\frac{\Delta}{2}\Big)^{2}\frac{\sin^{2}(\sqrt{D_{n+k}}t)}{D_{n+k}}
\Big], \nonumber \\
S_{2}(k)
&=&
\sum_{n=0}^{\infty}
\frac{|\alpha|^{2n}}{n!}
\frac{\sin^{2}(\sqrt{D_{n+k}}t)}{D_{n+k}}.
\label{S-1-2-k-t}
\end{eqnarray}

\section{\label{section-perturbative-expansion-rho-01}The perturbative expansion
of $\rho_{01}$}
Like in Eqs.~(\ref{probability-excited-state-atom-1}), (\ref{probability-excited-state-atom-2}), and (\ref{probability-excited-state-atom-3}),
we can formulate $\rho_{01}$ as
\begin{equation}
\rho_{01}
=
\sum_{n=0}^{\infty}\frac{\theta(\beta)^{n}}{n!}
\rho_{01}^{(n)}(\Theta;t),
\label{rho-01-1}
\end{equation}
\begin{equation}
\rho_{01}^{(n)}(\Theta;t)
=
\sin^{2}\Theta \rho_{01,1}^{(n)}(t)
+
\cos^{2}\Theta \rho_{01,2}^{(n)}(t),
\label{rho-01-2}
\end{equation}
\begin{eqnarray}
\rho_{01,1}^{(0)}(t)
&=&
{\cal P}^{(0)}(u^{\dagger}_{00}u_{10}),
\quad
\rho_{01,2}^{(0)}(t)
=
{\cal P}^{(0)}(u^{\dagger}_{01}u_{11}), \nonumber \\
\rho_{01,1}^{(1)}(t)
&=&
{\cal P}^{(1)}(u^{\dagger}_{00}u_{10}),
\quad
\rho_{01,2}^{(1)}(t)
=
{\cal P}^{(1)}(u^{\dagger}_{01}u_{11}), \nonumber \\
\rho_{01,1}^{(2)}(t)
&=&
{\cal P}^{(2)}(u^{\dagger}_{00}u_{10}),
\quad
\rho_{01,2}^{(2)}(t)
=
{\cal P}^{(2)}(u^{\dagger}_{01}u_{11}). \nonumber \\
\label{rho-01-3}
\end{eqnarray}

From slightly tough calculations, we obtain
\begin{equation}
\rho_{01,j}^{(0)}(t)
=
\tilde{S}_{j,0}(t)
\quad
\mbox{for $j=1,2$}
\end{equation}
\begin{eqnarray}
\rho_{01,j}^{(1)}(t)
&=&
-2|\alpha|^{2}\tilde{S}_{j,0}(t)
+
\alpha^{*}\tilde{S}_{j,1}(t)
+
\alpha\tilde{S}_{j,2}(t) \nonumber \\
&&
\mbox{for $j=1,2$},
\end{eqnarray}
\begin{eqnarray}
\rho_{01,j}^{(2)}(t)
&=&
2(-1-|\alpha|^{2}+2|\alpha|^{4})\tilde{S}_{j,0}(t) \nonumber \\
&&
-
(1+4|\alpha|^{2})\alpha^{*}\tilde{S}_{j,1}(t)
-
(1+4|\alpha|^{2})\alpha\tilde{S}_{j,2}(t) \nonumber \\
&&
+
\alpha^{*2}\tilde{S}_{j,3}(t)
+
\alpha^{2}\tilde{S}_{j,4}(t) \nonumber \\
&&
+
2(1+|\alpha|^{2})\tilde{S}_{j,5}(t)
\quad
\mbox{for $j=1,2$},
\end{eqnarray}
\begin{eqnarray}
\tilde{S}_{1,0}(t)
&=&
{}_{\mbox{\scriptsize B}}\langle\alpha|
u^{\dagger}_{00}u_{10}
|\alpha\rangle_{\mbox{\scriptsize B}} \nonumber \\
&=&
-ig
e^{-|\alpha|^{2}}
\sum_{n=0}^{\infty}
\frac{|\alpha|^{2n}\alpha^{*l}}{n!} \nonumber \\
&& \times
A(n+l)B'(n+l),
\end{eqnarray}
\begin{eqnarray}
\tilde{S}_{1,1}(t)
&=&
{}_{\mbox{\scriptsize B}}\langle\alpha|
au^{\dagger}_{00}u_{10}
|\alpha\rangle_{\mbox{\scriptsize B}} \nonumber \\
&=&
-ig
e^{-|\alpha|^{2}}
\sum_{n=0}^{\infty}
\frac{(n+l)|\alpha|^{2n}\alpha^{*(l-1)}}{n!} \nonumber \\
&& \times
A(n+l)B'(n+l),
\end{eqnarray}
\begin{eqnarray}
\tilde{S}_{1,2}(t)
&=&
{}_{\mbox{\scriptsize B}}\langle\alpha|
u^{\dagger}_{00}u_{10}a^{\dagger}
|\alpha\rangle_{\mbox{\scriptsize B}} \nonumber \\
&=&
-ig
e^{-|\alpha|^{2}}
\sum_{n=0}^{\infty}
\frac{|\alpha|^{2n}\alpha^{*(l+1)}}{n!} \nonumber \\
&& \times
A(n+l+1)B'(n+l+1),
\end{eqnarray}
\begin{eqnarray}
\tilde{S}_{1,3}(t)
&=&
{}_{\mbox{\scriptsize B}}\langle\alpha|
a^{2}u^{\dagger}_{00}u_{10}
|\alpha\rangle_{\mbox{\scriptsize B}} \nonumber \\
&=&
-ig
e^{-|\alpha|^{2}}
\sum_{n=0}^{\infty}
u(n+2-l) \nonumber \\
&& \times
\frac{(n+1)(n+2)|\alpha|^{2(n+2-l)}\alpha^{*(l-2)}}{(n+2-l)!} \nonumber \\
&& \times
A(n+2)B'(n+2),
\end{eqnarray}
\begin{eqnarray}
\tilde{S}_{1,4}(t)
&=&
{}_{\mbox{\scriptsize B}}\langle\alpha|
u^{\dagger}_{00}u_{10}a^{\dagger 2}
|\alpha\rangle_{\mbox{\scriptsize B}} \nonumber \\
&=&
-ig
e^{-|\alpha|^{2}}
\sum_{n=0}^{\infty}
u(n-2-l)
\frac{|\alpha|^{2(n-l-2)}\alpha^{*(2+l)}}{(n-l-2)!} \nonumber \\
&& \times
A(n)B'(n),
\end{eqnarray}
\begin{eqnarray}
\tilde{S}_{1,5}(t)
&=&
{}_{\mbox{\scriptsize B}}\langle\alpha|
au^{\dagger}_{00}u_{10}a^{\dagger}
|\alpha\rangle_{\mbox{\scriptsize B}} \nonumber \\
&=&
-ig
e^{-|\alpha|^{2}}
\sum_{n=0}^{\infty}
u(n-l)
\frac{(n+1)|\alpha|^{2(n-l)}\alpha^{*(l)}}{(n-l)!} \nonumber \\
&& \times
A(n+1)B'(n+1),
\end{eqnarray}
\begin{eqnarray}
\tilde{S}_{2,0}(t)
&=&
{}_{\mbox{\scriptsize B}}\langle\alpha|
u^{\dagger}_{01}u_{11}
|\alpha\rangle_{\mbox{\scriptsize B}} \nonumber \\
&=&
ig
e^{-|\alpha|^{2}}
\sum_{n=0}^{\infty}
\frac{|\alpha|^{2n}\alpha^{*l}}{n!}
B(n)A'(n),
\end{eqnarray}
\begin{eqnarray}
\tilde{S}_{2,1}(t)
&=&
{}_{\mbox{\scriptsize B}}\langle\alpha|
au^{\dagger}_{01}u_{11}
|\alpha\rangle_{\mbox{\scriptsize B}} \nonumber \\
&=&
ig
e^{-|\alpha|^{2}}
\sum_{n=0}^{\infty}
\frac{(n+l)|\alpha|^{2n}\alpha^{*(l-1)}}{n!} \nonumber \\
&& \times
B(n)A'(n),
\end{eqnarray}
\begin{eqnarray}
\tilde{S}_{2,2}(t)
&=&
{}_{\mbox{\scriptsize B}}\langle\alpha|
u^{\dagger}_{01}u_{11}a^{\dagger}
|\alpha\rangle_{\mbox{\scriptsize B}} \nonumber \\
&=&
ig
e^{-|\alpha|^{2}}
\sum_{n=0}^{\infty}
\frac{|\alpha|^{2n}\alpha^{*(l+1)}}{n!} \nonumber \\
&& \times
B(n+1)A'(n+1),
\end{eqnarray}
\begin{eqnarray}
\tilde{S}_{2,3}(t)
&=&
{}_{\mbox{\scriptsize B}}\langle\alpha|
a^{2}u^{\dagger}_{01}u_{11}
|\alpha\rangle_{\mbox{\scriptsize B}} \nonumber \\
&=&
ig
e^{-|\alpha|^{2}}
\sum_{n=0}^{\infty}
u(n+2-l) \nonumber \\
&& \times
\frac{(n+1)(n+2)|\alpha|^{2(n+2-l)}\alpha^{*(l-2)}}{(n+2-l)!} \nonumber \\
&& \times
B(n+2-l)A'(n+2-l),
\end{eqnarray}
\begin{eqnarray}
\tilde{S}_{2,4}(t)
&=&
{}_{\mbox{\scriptsize B}}\langle\alpha|
u^{\dagger}_{01}u_{11}a^{\dagger 2}
|\alpha\rangle_{\mbox{\scriptsize B}} \nonumber \\
&=&
ig
e^{-|\alpha|^{2}}
\sum_{n=0}^{\infty}
\frac{|\alpha|^{2n}\alpha^{*(2+l)}}{n!} \nonumber \\
&& \times
B(n+2)A'(n+2),
\end{eqnarray}
\begin{eqnarray}
\tilde{S}_{2,5}(t)
&=&
{}_{\mbox{\scriptsize B}}\langle\alpha|
au^{\dagger}_{01}u_{11}a^{\dagger}
|\alpha\rangle_{\mbox{\scriptsize B}} \nonumber \\
&=&
ig
e^{-|\alpha|^{2}}
\sum_{n=0}^{\infty}
\frac{(n+l+1)|\alpha|^{2n}\alpha^{*(l)}}{n!} \nonumber \\
&& \times
B(n+1)A'(n+1),
\end{eqnarray}
where
\begin{eqnarray}
A(n)
&=&
\cos(\sqrt{D_{n}}t)-i\frac{\Delta}{2}\frac{\sin(\sqrt{D_{n}}t)}{\sqrt{D_{n}}}, \nonumber \\
A'(n)
&=&
\cos(\sqrt{D'_{n}}t)-i\frac{\Delta}{2}\frac{\sin(\sqrt{D'_{n}}t)}{\sqrt{D'_{n}}}, \nonumber \\
B(n)
&=&
\frac{\sin(\sqrt{D_{n}}t)}{\sqrt{D_{n}}}, \nonumber \\
B'(n)
&=&
\frac{\sin(\sqrt{D'_{n}}t)}{\sqrt{D'_{n}}},
\end{eqnarray}
\begin{equation}
D'_{n}
=
\left\{
\begin{array}{ll}
(\Delta/2)^{2}
+
g^{2}
\prod_{k=1}^{l}(n-k+1)
&
\mbox{for $n\geq l$,} \\
(\Delta/2)^{2}
&
\mbox{for $n\leq l-1$,} \\
\end{array}
\right.
\end{equation}
\begin{eqnarray}
A'(n)
&=&
1-i(\Delta/2)t
\quad
\mbox{for $n\leq l-1$ and $\Delta=0$}, \nonumber \\
B'(n)
&=&
t
\quad
\mbox{for $n\leq l-1$ and $\Delta=0$,}
\end{eqnarray}
\begin{equation}
u(n)
=
\left\{
\begin{array}{ll}
1 & \mbox{for $n\geq 0$}, \\
0 & \mbox{for $n\leq -1$}, \\
\end{array}
\right.
\label{definition-u-n}
\end{equation}
and $D'_{n}$ is an eigenvalue of $D'$ for the number state of the photons $|n\rangle_{\mbox{\scriptsize P}}$.

\end{document}